\documentclass[12pt]{iopart}
\usepackage{graphicx}

\begin{document}

\title[BEG network]{\bf Thermodynamics of fully connected
                        Blume-Emery-Griffiths neural networks}
\author{D. Boll\'e and T. Verbeiren}
\address{Instituut voor Theoretische Fysica, KU Leuven, B-3001 Leuven, Belgium}
\ead{desire.bolle@fys.kuleuven.ac.be, toni.verbeiren@fys.kuleuven.ac.be}
\begin{abstract}

The thermodynamic and retrieval properties of fully connected
Blume-Emery-Griffiths networks, storing ternary patterns, are studied using
replica mean-field theory. Capacity-temperature phase diagrams are derived
for several values of the pattern activity. It is found that the retrieval
phase is the largest in comparison with other three-state neuron models.
Furthermore, the meaning and stability of the so-called quadrupolar phase is
discussed as a function of both the temperature and the pattern activity.
Where appropriate, the results are compared with the diluted 
version of the model.
\end{abstract}

\maketitle

\section{Introduction}

It has been argued recently that an optimal Hamiltonian, guaranteeing the
best retrieval properties for neural networks with multistate neurons can
be found by maximizing the mutual information content of the network
\cite{DK00,BV02}. In this way, for two-state neurons the well-known
Hopfield model is recovered \cite{H82}, for three-state neurons a
Blume-Emery-Griffiths (BEG) type spin-glass model is obtained
 \cite{BEG71}-\cite{ACN00} (and references therein).
In the discussion of the extremely diluted asymmetric version of
this model \cite{DK00}, which has an exactly solvable dynamics because
there are no feedback spin correlations \cite{BKS90}, a new phase appeared,
the so-called quadrupolar phase, that could yield new retrieval information.
The time evolution of the order parameter characterizing this phase together
with the stability properties of this phase in the extremely diluted
model have been studied \cite{BDEKT02}. Furthermore, the
zero-temperature
parallel dynamics of the fully connected architecture taking into account
all feedback correlations has been solved recently \cite{BBS02}. No
quadrupolar phase has been found in this case.

A complete study of both the thermodynamic and retrieval properties of a
fully connected architecture governed by such a BEG spin-glass Hamiltonian
is not yet given in the literature. Although some preliminary results about
the retrieval quality of this model for uniformly distributed patterns and
non-zero temperatures have been presented in \cite{BV02}, one would
like to have a complete temperature-capacity phase diagram as a function of
the  pattern activity. Furthermore, one would also like to solve the question
posed in \cite{BV02} about the stability of the quadrupolar phase.
Finally, one would like to know in more detail how the retrieval quality of
this model compares with other three-state neuron models. To fill these gaps
is the purpose of the present work.

The main results obtained are the following. Using replica-symmetric
mean-field theory it is  shown that the retrieval phase is systematically
larger than the one of the $3$-state neuron models known in the literature.
The critical capacity of the BEG neural network is about two times
bigger than the one of the $3$-state neuron Ising model \cite{BRS94}. 
The region of thermodynamic stability of the retrieval states is much 
larger than the one for the $3$-Ising model and, interestingly, even
slightly bigger than the corresponding region for the Hopfield model.  
Next, it is found that the quadrupolar phase is not a
stable solution for low temperatures but can become stable at high
temperatures for suitable choices of the network parameters. The physical
meaning of this is discussed. Finally, by calculating the zero-temperature
entropy we expect that, for uniformly distributed patterns, replica-symmetry
breaking is of the same order as the breaking in the Hopfield model.

The rest of the paper is organised as follows.  The model is introduced from
a dynamic point of view in section \ref{sect:model}. Section \ref{sect:calc}
presents the replica-symmetric mean-field approximation and obtains
the fixed-point equations for the relevant order parameters.  In Section
\ref{sect:results} these equations are studied in detail for arbitrary
temperatures. In particular, a temperature-activity phase diagram for low
loading and temperature-capacity phase diagrams for finite loading and
several pattern activities are obtained. Also the specific thermodynamic
properties are discussed. Where appropriate, the results are compared with
the diluted version of the model. Concluding remarks are given
in section \ref{sect:conclusion}. Finally, the appendix  contains the explicit
form of the fixed-point equations.

\section{The model}  \label{sect:model}

Consider a network of N neurons, $\{\sigma_i\}$, $i = 1, \dots N$, which take
values out of the set $\{-1, 0, 1\}$. In this network we want to store
$p=\alpha N$ patterns, $\{\xi_i^\mu\}$, $i = 1, \dots N$ and $\mu = 1, \dots p$.
They are supposed to be independent identically distributed random variables
(iidrv) with respect to $i$ and $\mu$ drawn from a probability distribution
given by
\begin{equation}
\rho(\xi_i^\mu) = a \, \delta(1 - (\xi_i^\mu)^2)
                     + (1-a) \, \delta(\xi^\mu_i)\, ,
\end{equation}
with $a$ the pattern activity, viz.
\begin{equation}
\lim _{N \to \infty} \frac{1}{N} \sum_i (\xi_i^\mu)^2 = a \ .
\end{equation}

The neurons are updated asynchronously according to the transition
probability
\begin{equation}
\mbox{Pr}\left(\sigma'_i=s\in \{-1,0,1\}|\{\sigma_i\} \right)=
     \frac{\exp[-\beta \epsilon_i(s|\{\sigma_i\})]}
          {\displaystyle{\sum_{s \in \{-1,0,1\}} 
	            \exp[-\beta \epsilon_i(s|\{\sigma_i\})] }}
\end{equation}
with $\beta$ the inverse temperature and $\epsilon_i(s|\{\sigma_i\})$ an
effective single site energy function given by \cite{DK00}
\begin{equation}
\epsilon_i(s|\{\sigma_i\})= 
  -h_i(\{\sigma_i\})s -\theta_i(\{\sigma_i\}) s^2 \, ,
           \quad s \in \{-1,0,1\}
\end{equation}
where the random local fields are defined by
\begin{equation}
h_i=
  \sum_{j=1}^N J_{ij} \sigma_j \, ,
     \qquad \theta_i=\sum_{j=1}^N K_{ij} \sigma_j^2 \, .
\end{equation}
The coefficients in these local fields are determined via the Hebb rule
\begin{equation}
\fl J_{ij} = \frac{1}{a^2 N} \sum_{\mu=1}^p \xi^\mu_i \xi^\mu_j \, ,
         \quad
K_{ij}=\frac{1}{a^2(1-a)^2 N} \sum_{\mu=1}^p \eta^\mu_i \eta^\mu_j \, ,
         \quad   \eta^\mu_i=((\xi^\mu_i)^2 - a) \ .
\end{equation}
For zero temperature, the dynamical rule becomes
\begin{equation}
\sigma'_i= \mbox{sign}(h_i(\{\sigma_i\})) \,
          \Theta(|h_i(\{\sigma_i\})|+\theta_i(\{\sigma_i\})) \, .
\end{equation}

The long-time behaviour of this network is governed by the Hamiltonian
 \cite{DK00, BV02}
\begin{equation}
H = - \frac{1}{2} \sum_{i \neq j} J_{ij} \sigma_i \sigma_j
    - \frac{1}{2} \sum_{i \neq j} K_{ij} \sigma_i^2 \sigma_j^2  
    \label{ha1} \, .
\end{equation}
Since we want to compare this model with the $3$-Ising model and we
want to be able to change the relative importance of the two terms we rewrite
the Hamiltonian as
\begin{equation}
H = - \frac{A}{2} \sum_{i \neq j} \tilde J_{ij} \sigma_i \sigma_j
    - \frac{B}{2} \sum_{i \neq j} \tilde K_{ij} \sigma_i^2 \sigma_j^2  
    \, ,
\end{equation}
with
\begin{equation}
   \tilde J_{ij}= a J_{ij} \, , \quad  \tilde K_{ij}= a(1-a) K_{ij} \, .
\end{equation}
For
\begin{equation} \label{BEGscaling}
A = \frac{1}{a}  \, , \quad  B = \frac{1} {a(1-a)}
\end{equation}
we trivially recover the model above.
When we now take $K_{ij} = b \delta_{ij}$ and $A=B=1$
we obtain the $3$-state Ising model \cite{BRS94}. Finally, we find back
the Hopfield model by taking first $B=0$ and then $a=1$, again with $A=1$.

\section{Replica-symmetric mean field theory}   \label{sect:calc}

We apply the standard replica technique \cite{MPV78} in order to calculate the
free energy of the model.  Within the replica-symmetry approximation and for
a finite number, $s$, of condensed patterns, we obtain
\begin{eqnarray} \label{free_energy}
 f(\beta) =  &\frac{1}{2} \sum_{\nu=1}^s
       \left( a A\, m_\nu^2 + a(1-a) B \, l_\nu^2
       \right)
  + \frac{\alpha}{2\beta} \log (1-\chi)
  + \frac{\alpha}{2\beta} \log (1-\phi) \nonumber \\
  & + \frac{\alpha}{2\beta} \frac{\chi}{1-\chi}
   + \frac{\alpha}{2\beta} \frac{\phi}{1-\phi}
  + \frac{\alpha}{2} \frac{A q_1 \chi}{(1-\chi)^2}
  + \frac{\alpha}{2} \frac{B p_1 \phi}{(1-\phi)^2} \nonumber \\
  & - \frac{1}{\beta} \left\langle \int Ds Dt \ln
       \mbox{Tr}_\sigma
       \exp \left ( \beta \tilde H \right )
    \right\rangle_{\{\xi^\nu\}} \ ,
\end{eqnarray}
with the effective Hamiltonian $\tilde H$ given by
\begin{equation}
\fl
\tilde H  = A \sigma
 \left[
   \sum_\nu m_\nu \xi^\nu
        + \sqrt{\alpha r}s
 \right]
+ B \sigma^2
 \left[
   \sum_\nu l_\nu \eta^\nu + \sqrt{\alpha u} t
 \right]
+ \frac{\alpha}{2} \frac{  A \, \chi}{1-\chi} \, \sigma^2
+ \frac{\alpha}{2} \frac{  B \, \phi}{1-\phi} \, \sigma^2
\ ,
\end{equation}
and where $Ds$ and $Dt$ are Gaussian measures,
 $Ds=ds (2\pi)^{-1/2}\exp(-s^2/2)$. Furthermore
\begin{equation}
\fl
\chi = A \beta ( q_0 - q_1 )\, , \quad  \phi = B \beta ( p_0 - p_1)\, ,
   \quad r = \frac{q_1}{(1-\chi)^2}\, ,
   \quad u = \frac{p_1}{(1-\phi)^2} \, .
\end{equation}
In these expressions the relevant order parameters are
\begin{eqnarray}
m_\nu & = &
\frac{1}{a}
\left\langle  \xi^\nu \int \!  Ds Dt \
  \left\langle \sigma \right\rangle_\beta
\right\rangle_{\{\xi^\nu\}}
\ , \label{sp:m}
\\
l_\nu & = &
\frac{1}{a(1-a)}
\left\langle \eta^\nu \int \! Ds Dt \  \left\langle \sigma^2
\right\rangle_\beta
\right\rangle_{\{\xi^\nu\}}
\ , \label{sp:l}
\\
q_0 &  = &p_0 =
\left\langle
\int \! Ds Dt \  \left\langle \sigma^2 \right\rangle_\beta
\right\rangle_{\{\xi^\nu\}}
\ , \label{sp:q_0}
\\
q_1 & = &
\left\langle
\int \! Ds Dt \  {\left\langle \sigma
\right\rangle}_\beta^2
\right\rangle_{\{\xi^\nu\}}
\ , \label{sp:q_1}
\\
p_1 & = &
\left\langle
\int \! Ds Dt \  {\left\langle
\sigma^2 \right\rangle}^2_\beta
\right\rangle_{\{\xi^\nu\}}
\ , \label{sp:p_1}
\end{eqnarray}
where $\left\langle \cdot \right\rangle_\beta$ represents the thermal average
with respect to the effective Hamiltonian $\tilde H$.
In the sequel we take  only one condensed pattern such that the index
 $\nu$ can be dropped.

The parameter $m$ is the usual overlap between the condensed pattern and
the network state, while $l$ is related to the activity overlap, i.e.,
$\langle \sigma^2 \xi^2 \rangle$. In
\cite{BDEKT02} it has been called the fluctuation overlap between the
binary state variables $\sigma_i^2$ and $\eta_i^2$. Furthermore, $q_0$
is the activity of the neurons and $q_1$ and $p_1$ are the
Edwards-Anderson order parameters with their conjugate variables $r$,
respectively $u$. Finally, $\chi$ and $\phi$ are the susceptibilities
proportional to the fluctuation of the $m$ overlap, respectively $l$
overlap.

We remark that the trace over the neurons and the average over the
patterns can be performed explicitly. The resulting expressions  are
written down in  \ref{app:1}.

\section{Thermodynamic and retrieval properties}    \label{sect:results}

In this section, we study the thermodynamic and retrieval properties of
the fully connected BEG network by numerically solving the fixed-point
equations (\ref{sp:m})-(\ref{sp:p_1}) for one condensed pattern.
Depending on the temperature $T$ and on the system parameters $\alpha$
and $a$ we recognize the following phases in terms of the order
parameters.

There is a  retrieval phase $R$ ($m>0, l>0, q_1 >0$) characterized by 
positive $m$ and $l$ and a quadrupolar phase $Q$ ($m=0, l>0, q_1=0 $)
where only  $l$ is non-zero, meaning that the active neurons
($\pm 1$) coincide with the active patterns but the signs are not
correlated. This implies that this phase also carries some retrieval
information. From the fact that $q_1=0$ in this
phase we know that the spins are not frozen, so that we expect to find
this phase only for high enough temperatures. Furthermore, there is the
spin-glass phase $S$ ($m=0, l=0, q_1 >0 $) and the paramagnetic phase
$P$ ($m=0, l=0, q_1 =0 $). We first look at low loading $\alpha=0$.

\subsection{Low loading}

\begin{figure}[hbt]
\begin{center}

\includegraphics[width=.6\textwidth,clip=]{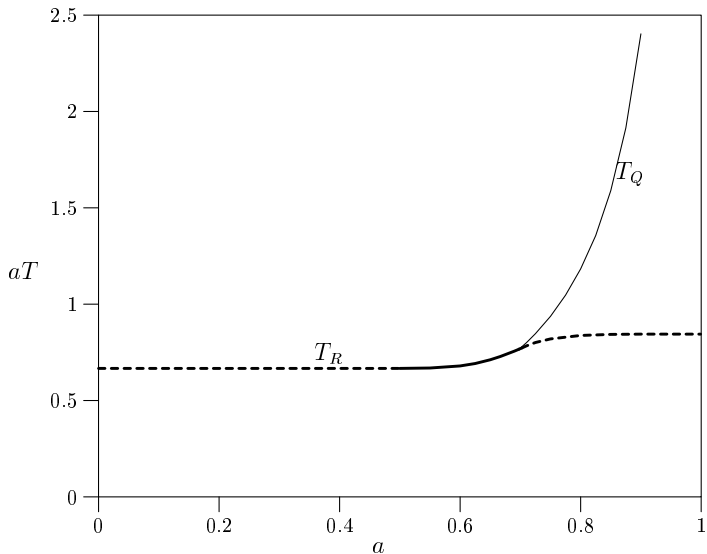}

\caption{The BEG phase diagram for $\alpha=0$ :  $aT$ as a function of
$a$. Dashed (full) lines denote continuous (discontinuous)
transitions.}

\label{pd:alpha0}

\end{center}
\end{figure}

For $\alpha=0$, the fixed-point equations simplify a lot because all the
integrations (see \ref{app:1}) drop out. This allows us to carefully
study the quadrupolar state as a function of the pattern activity $a$ since
it turns out that the effect of the quadrupolar phase is strongest for a small
loading capacity and high temperatures. A temperature-activity phase is
presented in figure \ref{pd:alpha0}. A dashed (full) line corresponds to 
a continuous (discontinuous) transition.

Below $a=1/2$, there is a continuous transition at $aT_R=2/3$ from the
retrieval phase at low $T$ to the  paramagnetic phase at high $T$.  This is
similar to the $3$-Ising model where it occurs for all values of $a$
\cite{BRS94}. At $a=1/2$ the transition becomes discontinuous
and, up to $a = 0.698$, the only phases present are $R$ and $P$. The
quadrupolar phase starts to appear at $a = 0.698$ and $aT=0.767$,
and beyond that point it keeps growing for increasing $a$.  The transition
$R-Q$ remains discontinuous up to $a = 0.708$ and $aT=0.78$ . For
bigger values of $a$, it remains continuous
and ends at $aT=(1-(2e)^{-1})^{-1}=0.64$ for $a=1$.
We remark that this phase diagram is the same, up to some slightly
different numerical values for the transition points, as the
corresponding one for
the extremely diluted model \cite{BDEKT02}, confirming the fact that for low
loading the architecture is not important.

\subsection{Finite loading}

\begin{figure}[hb]
  \begin{center} 
  \includegraphics[width=.6\textwidth]{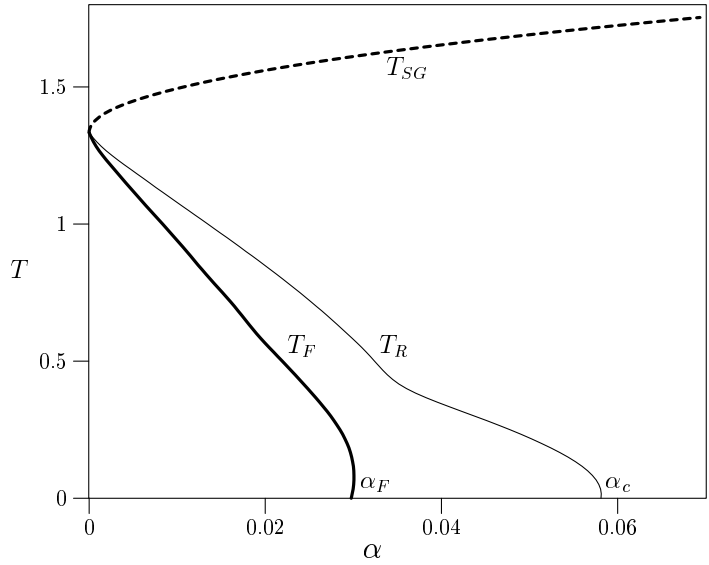}
  
  \caption{
    The BEG $\alpha-T$ phase diagram for $a=1/2$.
    The meaning of the lines is explained in the text.
    \label{pd:a0_50}
  }
  \end{center}
\end{figure}

\begin{figure}[b]
  \begin{center} 
   \includegraphics[width=.6\textwidth]{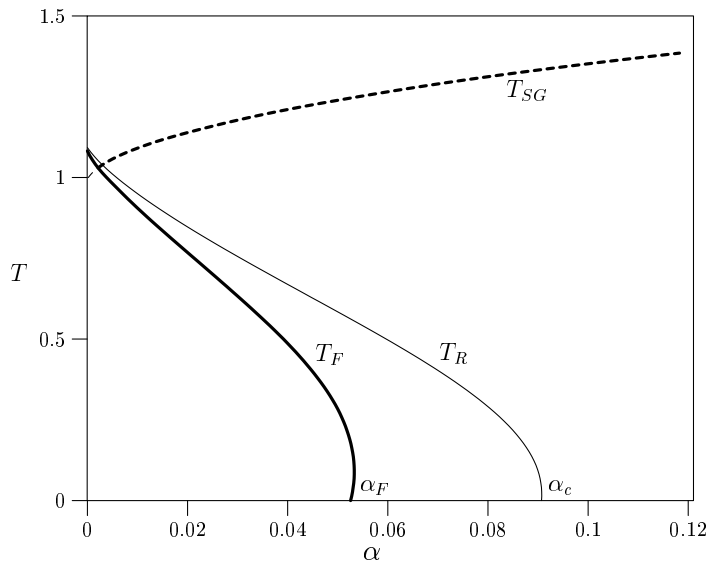}
   
   \caption{
    The BEG $\alpha-T$ phase diagram for $a=2/3$.  
    The meaning of the lines is explained in the text
    \label{pd:a2__3}
  }
  \end{center}
\end{figure}

\begin{figure}[b]
  
  \begin{center} 
  \includegraphics[width=.6\textwidth]{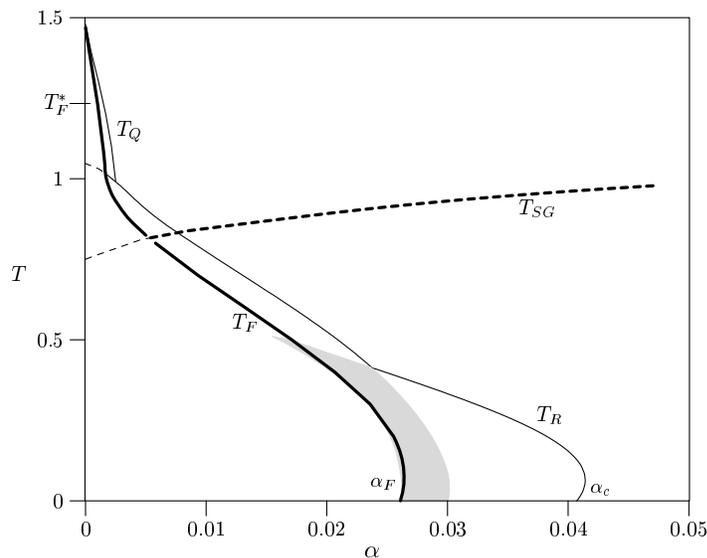}
  
  \caption{
    The $\alpha-T$ BEG phase diagram for $a=.8$.
    The meaning of the lines is explained in the text.
    \label{pd:a0_80}
  }
  \end{center}
\end{figure}

Next, in figures \ref{pd:a0_50}, \ref{pd:a2__3} and \ref{pd:a0_80} we
present the temperature-capacity phase diagrams for three typical values of
the pattern activity, $a=1/2$, $a=2/3$ (uniformly distributed patterns) and
$a=0.8$.  As before, a dashed line corresponds to a continuous transition,
while a full line corresponds to a discontinuous transition (in all order
parameters).

We start with some general observations. Below the line $T_R$ retrieval
states occur.  The curve
$T_F$ represents the thermodynamic transition between retrieval
states and spin-glass states. Hence, below $T_F$ the retrieval states
are global minima of the free energy while above this line the 
spin-glass states are. Thermodynamic transitions are shown as thick lines.
The line $T_{SG}$ denotes the transition from the spin-glass to the
paramagnetic phase.

In all phase diagrams we notice some reentrance, i.e., $\alpha_c$ at some
finite $T$ is larger than $\alpha_c$ at $T=0$. This effect is
well-known from the literature (see, e.g., \cite{WS91}, \cite{NC92}) and
signals the breaking of replica symmetry. In order to have an idea about
the size of this breaking compared with other models, we calculate the
entropy at zero temperature \cite{AGS87}:
\begin{equation}
S = -\frac{\alpha}{2}
 \left [
   \ln(1-\chi) + \ln(1-\phi)
     + \frac{\chi}{1-\chi} + \frac{\phi}{1-\phi}
 \right ] \ .
\end{equation}
We find that this entropy is indeed negative but small. For uniform patterns,
e.g., the entropy of the retrieval state at $\alpha_c$ is, 
$S(\alpha_c)=-0.0017$, which is
of the same order of magnitude as the one for the Hopfield model, i.e.,
$S(\alpha_c)=-0.0014$, suggesting that the breaking is comparable. 
 Moreover, the 
zero temperature entropy becomes more negative for increasing $a$, e.g.,
$S(\alpha_c)$ is $-0.0010$, $-0,0017$ and $-0.031$ for $a=1/2$, $a=2/3$ and 
$a=0.8$ respectively, which might suggest that the breaking becomes
larger.

Next, we look at the different phase diagrams in more detail. In the case of
$a=1/2$ shown in \fref{pd:a0_50}, the diagram quantitatively resembles the one
for the $3$-Ising model \cite{BRS94}. At high temperatures there is the
continuous transition from the disordered paramagnetic phase to the spin-glass
phase. When crossing the curve $T_R$ retrieval states show up as local
minima of the free energy. At these points the overlap with the embedded
patterns jumps from zero to a finite macroscopic value. In comparison
with the $3$-Ising model there is a small kink in the line $T_R$. When
lowering $T$ further the retrieval states become global minima of the free
energy. This happens along the curve $T_F$ and this thermodynamic
transition is first order.
The value for the critical capacity at zero temperature is
$\alpha_c = 0.058$, while the thermodynamic transition point is
$\alpha_F = 0.030$.

For uniform patterns (see \fref{pd:a2__3}) several transition curves
bordering the different phases discussed above show up. Here we note
that the critical curves $T_{SG}$ and $T_R$ end in different temperature
points at $\alpha=0$ giving rise to a `crossover' region for small
$\alpha$ as it occurs in the Potts model \cite{BDH92}. This is related
with the fact that for $\alpha=0$ this model has a discontinuous
transition at $T_F$. In this crossover region the retrieval states
(global minima below $T_F$) and the paramagnetic states (local minima
below $T_F$) coexist.
Comparing these results with those found for the $3$-Ising model
\cite{BRS94},  we see that the $\alpha_c=0.091$ found here is almost double 
of the
critical capacity of the latter, $\alpha_c=0.046$. We recall that the
last number is obtained in the case of an optimal 
choice for the model parameter $b$, i.e. $b=1/2$, which makes the Hamming 
distance  minimal \cite{BRS94}. Furthermore, the region of thermodynamic 
stability
for the retrieval states is about four times bigger. Compared with the
Hopfield  model, we notice that $\alpha_c$ is smaller in the BEG model,
$0.091$ versus $0.13$, but $\alpha_F$ is larger, $0.053$ versus $0.051$.
So a bigger number of the retrieval states in the BEG network are global
minima of the free energy.

Figure \ref{pd:a0_80} shows the phase diagram for $a=0.8$.  We immediately
remark  that the structure of the phase diagram turns out to be more 
complicated, as expected.
The major difference with the foregoing phase diagrams is the presence of
the quadrupolar phase for high temperatures. Indeed, from the discussion in
section 4.1 (see figure
\ref{pd:alpha0}) we know that this quadrupolar phase shows up from an 
activity $a=0.698$ onwards.
As a consequence, the cross-over or coexistence  region is larger than the
one for the $a=2/3$ phase diagram.
The transition from the $Q$ phase to the $P$ phase is discontinuous
and by comparing the relevant free energies of the different phases we
find the thermodynamic transition line $T_F^*$ below which the
quadrupolar states are global minima. We see that $T_F^*$ joins nicely
with $T_F$, as it should.  The transition from the $R$ to the $Q$ phase is
continuous for small $\alpha$ up to $\alpha = 0.0023$ and discontinuous
beyond that value.
Finally, we notice that there is a larger reentrance suggesting a
stronger replica-symmetry breaking,  which is consistent with the
fact that the zero-temperature entropy is more negative in this case, as
mentioned above. 

\begin{figure}[b]
  
  \begin{center} 
  \includegraphics[width=.6\textwidth]{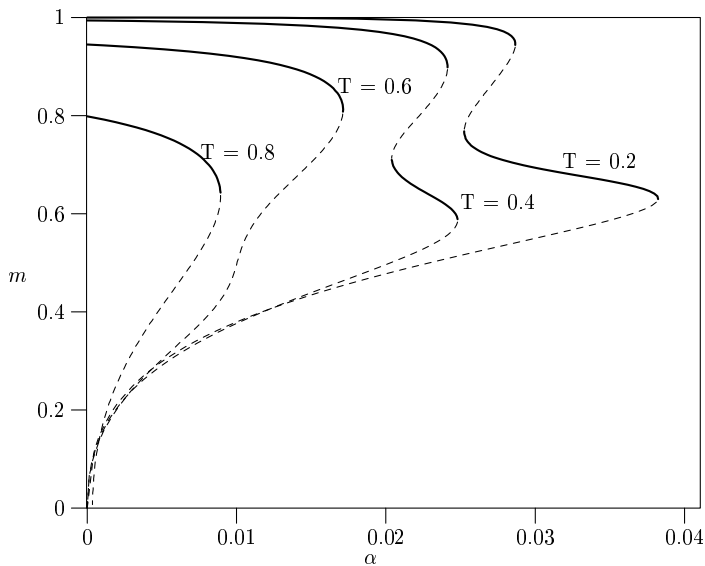}

  \caption{
    The overlap $m$ as a function of $\alpha$ for $T=0.2, 0.4, 0.6, 0.8$.  
    Full
    (dashed) lines denote stable (unstable) solutions.
    \label{fig:m-alpha}
  }
  \end{center}
\end{figure}

The quadrupolar phase is situated in the high temperature region and we
can understand the physics behind it in the following way.  The spin-glass
order parameter $q_1$ is zero, meaning that the $\pm 1$ spins are not
frozen and as a consequence $m$ can be zero.  The fact that $l$ is not
zero practically means that the spins can flip freely between $\pm 1$
but the probability that they jump to 0 or vice versa becomes very small. 
This effect arises from $a>1/2$
onwards when the ratio between the second and the first term in the 
Hamiltonian starts increasing as $(1-a)^{-1}$. It implies that the
information content of the system is non-zero in this phase. A practical
example may be
in pattern recognition where, looking at black and white pictures on a grey
background, this phase would tell us the exact location of the picture
with respect to the background without finding the details of the picture
itself.

Returning to figure \ref{pd:a0_80} we remark that in the shaded region
two retrieval states
coexist. A similar behaviour has also been seen in other multi-state
networks, e.g., the fully connected Potts model \cite{BDH92}. Their
presence can be understood by studying the overlap order parameter $m$
as a function of $\alpha$. This is shown in figure \ref{fig:m-alpha} 
where the full line
denotes stable solutions, while the dashed line corresponds to unstable
solutions (saddle points).  For small $T$ two stable solutions occur. The
one with the smallest overlap vanishes for large $T$.

We end this section with a  technical remark. In studying the fixed-point
equations of this model, it turns out that a lot of solutions are in fact
saddle points and not minima of the free energy.  This can, of course, be
investigated by studying the local stability of the extrema of the free
energy. For large $a$ we find, e.g.,  that there may be more than 4 possible
solutions involving some kind of quadrupolar character ($m=0$, $l>0$).
Only the solution depicted in figure \ref{pd:a0_80} is a stable one (a real 
atractor).  This also answers the question posed in 
\cite{BV02} about the stability of the quadrupolar state at low temperatures.

\subsection{Coefficients $A$ and $B$  versus the critical capacity}

In defining the BEG neural network model, we have included the possibility of
varying the coefficients $A$ and $B$ in the Hamiltonian.  
The value of the coefficients $A$ and $B$ given by
(\ref{BEGscaling}) stems from \cite{DK00,BV02}.  They are obtained by 
optimizing the mutual information of the system. One could expect that this 
choice also improves other properties of the neural network, e.g., the 
basin of attraction, the critical capacity $\alpha_c$.

\begin{figure}[ht]
  \begin{center} 
   \includegraphics[width=.6\textwidth]{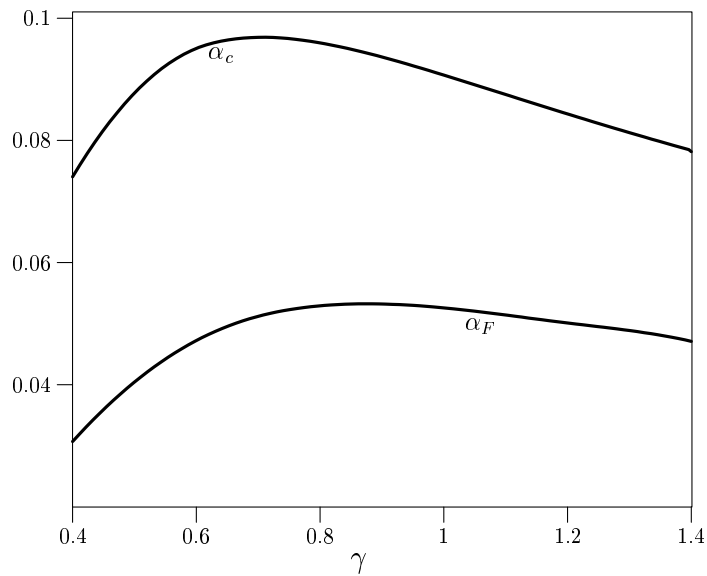}
   
   \caption{The capacities $\alpha_c$ and $\alpha_F$ as a function of 
   $\gamma = a(1-a)B$ for uniform patterns at $T=0$.
   \label{alpha_vs_B}
   }
  \end{center}
\end{figure}

Figure \ref{alpha_vs_B} shows $\alpha_c$ (dashed line) and $\alpha_R$ (full 
line) as a function of $\gamma=a(1-a)B$ with $A$ fixed at $A=1/a$ for 
uniform patterns at $T=0$. For $\gamma=1$, we recover the BEG neural network
as in (\ref{BEGscaling}) and studied above. It
turns out that the maximum in the critical capacity is located at
$\gamma=0.712$ with a corresponding value of $\alpha_c = 0.096$.
Also the maximum in the thermodynamic transition is located at a value
smaller than 1.  This does not agree with the expectation formulated above.

A reason for this is the approximation done in order to get the mean-field
Hamiltonian in \cite{DK00,BV02}.  At a certain point in its derivation 
the authors assume that $q_0 \sim a$. Consequently, the mutual information of
the network is optimized under this assumption.  Although this assumption is 
natural for having a complete match between the final state of the network 
and the condensed pattern, in general, it may  not be realized in a specific
model.
Furthermore, the fact that replica-symmetry breaking may be bigger for larger
$\alpha$, as is also suggested by the zero-temperature entropy
calculation, could be an extra reason why the maximum in $\alpha_c$ is further 
away from the  value $\gamma=1$ than is the maximum in $\alpha_F$. 
A further investigation is non-trivial and beyond the scope of the
present work.

\section{Concluding remarks}   \label{sect:conclusion}

We have considered both the thermodynamic and retrieval properties of
fully connected BEG networks. Fixed-point equations for the relevant
order parameters have been derived for
arbitrary temperatures in the replica-symmetric mean-field
approximation. An activity-temperature phase diagram for low loading has
been obtained. Near saturation, capacity-temperature phase diagrams have 
been discussed in detail for several values of the  activity of the 
three-state patterns. 

Compared with existing three-state
neuron models the retrieval region is larger and, e.g., for uniformly
distributed patterns the critical capacity at zero temperature is almost
two times bigger. Also the region of thermodynamic stability of the
retrieval states is much
enlarged and even larger than the one for the two-state Hopfield model.
A new information carrying phase, the quadrupolar phase, appears at larger
values of the activity in the high-temperature region of the phase
diagram and may extend the practical usefulness of this network, e.g, in
pattern recognition.

\ack

We would like to thank J. Busquets-Blanco, R. Erichsen jr., I. P\'erez-Castillo
and W. Theumann for fruitful discussions. One of the authors (DB) thanks
W. Theumann
for hospitality and financial support during a stay at the Instituto de F\'isica
of the Universidade Federal do Rio Grande do Sul, Porto Alegre, Brazil, where
part of this work has been done. This work has been supported in part
by the Fund of Scientific Research, Flanders-Belgium.

\section*{References}

\newpage

\appendix

\section{Explicit expressions for the fixed point equations} \label{app:1}

In this appendix we write down explicitly the fixed-point equations for the
order parameters of the BEG network. After performing the trace over the spins
and the average over the condensed patterns in equations
(\ref{sp:m})-(\ref{sp:p_1}) we obtain

\begin{eqnarray}  \label{sp:long}
\fl m = \int \! Ds Dt \ \mathbf{V}_\beta
         \left[
            \left( A m + A \sqrt{\alpha r}s \right),
            \left( B l(1-a) + B \sqrt{\alpha u} t + \frac{\alpha}{2} X \right)
         \right]\\
\fl l = \int \! Ds Dt \ \mathbf{W}_\beta
         \left[
            \left( A m + A \sqrt{\alpha r}s \right),
            \left( B l(1-a) + B \sqrt{\alpha u} t + \frac{\alpha}{2} X \right)
         \right] \nonumber\\
 \fl \hspace*{1cm}  \qquad - \int Ds Dt \ \mathbf{W}_\beta
         \left[
            \left( A \sqrt{\alpha r}s \right),
            \left(-B l a+ B \sqrt{\alpha u} t + \frac{\alpha}{2} X \right)
         \right]\\
\fl q_0  = p_0 = a \int \! Ds Dt \ \mathbf{W}_\beta
         \left[
            \left( A m + A \sqrt{\alpha r}s \right),
            \left( B l(1-a) + B \sqrt{\alpha u} t + \frac{\alpha}{2} X \right)
         \right] \nonumber\\
 \fl \hspace*{1cm}   \qquad + (1-a) \int Ds Dt \ \mathbf{W}_\beta
         \left[
            \left( A \sqrt{\alpha r}s \right),
            \left(-B l a + B \sqrt{\alpha u} t + \frac{\alpha}{2} X \right)
         \right]\\
\fl q_1 = a \int \! Ds Dt \ \left( \mathbf{V}_\beta
         \left[
            \left( A m + A \sqrt{\alpha r}s \right),
            \left( B l(1-a) + B \sqrt{\alpha u} t + \frac{\alpha}{2} X \right)
         \right] \right)^2 \nonumber\\
\fl \hspace*{1cm}    \qquad + (1-a) \int Ds Dt \ \left( \mathbf{V}_\beta
         \left[
            \left( A \sqrt{\alpha r}s \right),
            \left(-B l a + B \sqrt{\alpha u} t + \frac{\alpha}{2} X \right)
         \right] \right)^2\\
\fl p_1 = a \int \! Ds Dt \ \left( \mathbf{W}_\beta
         \left[
            \left( A m + A \sqrt{\alpha r}s \right),
            \left( B l(1-a) + B \sqrt{\alpha u} t + \frac{\alpha}{2} X \right)
         \right] \right)^2 \nonumber\\
\fl \hspace*{1cm}    \qquad + (1-a) \int Ds Dt \ \left(\mathbf{W}_\beta
         \left[
            \left( A \sqrt{\alpha r}s \right),
            \left(-B l a + B \sqrt{\alpha u} t + \frac{\alpha}{2} X \right)
         \right] \right)^2
\end{eqnarray}
with
\begin{equation}
\fl
\chi = A \beta ( q_0 - q_1 )\, , \quad  \phi = B \beta ( p_0 - p_1)\, ,
   \quad r = \frac{q_1}{(1-\chi)^2}\, ,
   \quad u = \frac{p_1}{(1-\phi)^2} 
\end{equation}
and
\begin{equation}
X  = A \frac{\chi}{1-\chi} + B \frac{\phi}{1-\phi} \, .
\end{equation}
The functions $\mathbf{V}_\beta$ and $\mathbf{W}_\beta$ are defined by
\begin{eqnarray}
\mathbf{V}_\beta(x,y)
  & = \frac
         {\sinh(\beta x)}
         {\frac{1}{2} \exp(-\beta y) + \cosh(\beta x)} \label{V} \, ,\\
\mathbf{W}_\beta(x,y) 
  & = \frac
         {\cosh(\beta x)}
         {\frac{1}{2} \exp(-\beta y) + \cosh(\beta x)} \label{W} \, ,
\end{eqnarray}
and reduce, for zero temperature, to
\begin{eqnarray}
\mathbf{V}_\infty(x,y)
  & = \mbox{sign}(x) \, \Theta(|x| + y) \ ,\\
\mathbf{W}_\infty(x,y)
  & = \Theta(|x| + y) \ .
\end{eqnarray}

\end{document}